\documentclass[preprintnumbers,twocolumn,amsfonts,amssymb,amsmath,floatfix,nofootinbib]{revtex4}
\usepackage[english]{babel}
\usepackage[T1]{fontenc}
\usepackage{psfrag}
\usepackage[dvips,final]{graphicx}
\usepackage{esint}

\newcommand{\be}{\begin{equation}}
\newcommand{\ee}{\end{equation}}

\newcommand{\etion}[1]{\ensuremath{\left\langle #1 \right\rangle}}

\newcommand{\dbar}{{\mathchar'26\mkern-11mu\mathrm{d}}}

\newcommand{\ds}{\mathrm{d}}

\newcommand{\Rcal}{\ensuremath{\mathcal{R}}}

\newcommand{\intfb}[1]{\ensuremath{\int\dbar^4k \,}}
\newcommand{\inttb}[1]{\ensuremath{\int\dbar^3k \,}}

\newcommand{\pdbd}[2]{\ensuremath{\frac{\partial #1}{\partial #2}}}

\newcommand{\Mpl}{M_\text{Pl}}
\newlength\rringshift 
\newcommand*\mathrring[1]{%
   \setbox0=\hbox{$#1$}%
   \dimen0 \wd0
   \advance\dimen0 -\rringshift
   \wd0 \dimen0
   \mathring{\copy0}%
   \kern-\wd0
   \kern \rringshift
   \mathring{%
     \phantom{\copy0}%
   }%
}
\newcommand{\tabfrac}[2]{%
	\setlength{\fboxrule}{0pt}%
	\fbox{$\frac{#1}{#2}$}%
}

\begin{document}
\title{Dark Matter via Many Copies of the Standard Model}
\author{Gia Dvali$^{1,2}$}
\email{georgi.dvali@cern.ch}
\author{Ignacy Sawicki$^{2}$}
\email{ignacy.sawicki@nyu.edu}
\author{Alexander Vikman$^{2}$}
\email{alexander.vikman@nyu.edu}
\affiliation{
${}^1\!$Theory Division, CERN,CH-1211 Geneva 23, Switzerland\\
${}^2\!$Center for Cosmology and Particle Physics, Department of Physics, NYU, New York,  NY 10003, USA}
\preprint{CERN-PH-TH/2009-027}

\begin{abstract}
We propose a cosmological scenario based on the assumption that the Standard Model possesses a large number of copies. It is demonstrated that baryons in the hidden copies of the standard model can naturally account for the dark matter. The right abundance of the hidden-sector baryons and the correct spectrum of density perturbations are simultaneously generated during modulated reheating. We show that for the natural values of inflaton coupling constants, dictated by unitarity, the dark-matter abundance is predicted to be proportional to the ratio of observed cosmological parameters: the square of the amplitude of cosmological perturbations and the baryon-to-photon number ratio.
\end{abstract}

\maketitle

\section{Introduction}

One of the major unsolved problems in cosmology is the understanding of the nature of dark matter (``DM''). The influence of its gravity can be seen not only in the local universe: in rotation curves of galaxies, gravitational lensing by them and the temperature of gas in clusters; but also in the form of the anisotropies in the cosmic microwave background. For an overview of known DM properties and manifestations see e.g. textbooks \cite{Mukhanov:2005sc, Weinberg}. All cosmological observations to date are consistent with the fact that the universe's energy budget comprises approximately 4\% baryonic matter, 21\% dark matter and 75\% dark energy. Astrophysical data imply that if dark matter interacts in any other way than through gravity, it does so extremely weakly, leaving no observational signature of this fact large enough to be detectable to date (the debate as to whether the recent PAMELA \cite{Adriani:2008zr} and ATIC \cite{Chang:2008zzr} signals are a result of dark-matter annihilation notwithstanding).

Indeed, some of the most stringent constraints on the collider physics of supersymmetry or extra dimensions arise as a result of the requirement to predict the appropriate dark-matter abundance. These extensions of the standard model (``SM'') are invoked, among others, to account for the hierarchy problem and contain within them natural candidates for dark matter.

Recently, a different solution to  the hierarchy problem was proposed in \cite{Dvali:2007hz, Dvali:2007wp}. Namely, it was realized that in any effective field theory with a large number $N$ of elementary particle species, the fundamental scale at which gravity becomes strong is lower than the Planck mass $\Mpl$ and is given by
  \be\label{bound}
        M_* \equiv \frac{\Mpl}{\sqrt{N}}.
  \ee
This scale serves as the cut-off for the low-energy effective theory. The generalization of the bound \eqref{bound} to quasi-de-Sitter inflationary vacua was given in \cite{Dvali:2008sy}. 
Notice, that the scale $M_*$ also puts the upper bound on the mass of the species of number $N$. Therefore, if there were to exist $N=10^{32}$ different species in our universe, this would place the scale beyond which gravity is strong at 1~TeV. In this way, the strong-gravity effects can potentially provide a mechanism which regulates the ultraviolet-sensitive radiative corrections to the mass of the Higgs. The existence of this strong-coupling scale was initially proved by considering the implications of models with a large number of species on the consistency of black-hole (``BH'') evaporation \cite{Dvali:2007hz, Dvali:2007wp}. Moreover, the same result has been justified by considering BH entanglement entropy \cite{Dvali:2008jb}  and quantum information theory \cite{Dvali:2008ec} and holography \cite{Dvali:2008ec,Horvat:2008ze}.  The possibility of the perturbative generation of the hierarchy \eqref{bound} between the Planck mass and fundamental gravity scale as a result of the renormalization of the gravitational coupling by the loops of $N$-species was suggested even earlier \cite{Dvali:2000xg,Veneziano:2001ah}. However, in contrast to the non-perturbative BH arguments, the latter's one-loop mechanism had to be based on naturalness assumptions, owing to the usual limitations of perturbation theory.

In this paper, we work within the above framework and suggest a mechanism which naturally produces the appropriate abundance of dark matter. We propose that there exist an additional, discrete permutation symmetry, $P(N)$, under which the standard model is charged, and therefore, there exist $N$ copies of the SM augmented by $N$ copies of the inflaton.

Our idea is that the matter we usually consider to be baryonic all belongs to the same species, while the dark matter consists of baryons of all the other copies of the standard model.
Current astrophysical observations constrain DM to be effectively a collisionless fluid that behaves as if it only interacted gravitationally. In our scenario, this property results from: (a) Low partial densities of each species except ours, which are achieved through the proposed inflationary and reheating mechanisms; (b) The suppression of any cross-couplings between different species compared to couplings within one sector, which---we demonstrate--- follows directly from unitarity.

The idea of dark matter's being stored in the form of the several hidden copies of baryons was put forward in \cite{ArkaniHamed:1999zg} in the context of extra-dimensional theories.  Since this mechanism was not tied to the solution of the hierarchy problem, the number of copies was relatively small, but otherwise unconstrained. Nor was there any specific connection predicted between the dark-matter density and the mechanism for the generation of cosmological perturbations. In the context of a permutation-symmetric standard model, this idea is also considered in a complementary work \cite{Dvali:2009dark}.

In addition, the possibility of mirror dark matter with a single SM mirror copy was also studied in \cite{Berezhiani:2000gw}. However, as opposed to the many-copy scenario, both sectors must contain additional DM particles and such a possibility is beyond our interest.

In other works, \cite{Adshead:2008gk,Watanabe:2007tf,Leblond:2008gg} inflation with a large number of species is also considered, however, with a different focus to this paper.

In our scenario, the large number of species leads to a number of qualitatively novel features:

Consistent inflation, without strong-gravity effects, can occur only below the cutoff $M_*$, given by Eq.~\eqref{bound}. In this case, it is natural \cite{Sawicki:2009} to implement a modulated-reheating scenario \cite{Dvali:2003em,Dvali:2003ar} to generate the appropriate amplitude for the power spectrum of cosmological perturbations.   In order to populate one of the species more than any other, the final stage of inflation should occur in one species only, so that reheating happens mostly within this preferred sector. However, a cross-coupling of the inflaton with the other species, which naturally appears at least owing to gravity, ensures that some of the inflaton decays occur into all the other copies. The cosmological perturbations generated in this scenario are guaranteed to be adiabatic if the modulator is common to all species.

It is important that the partial decay rates of the inflaton field into different species are strongly restricted by unitarity considerations. As a result, dark-matter abundance is predicted in terms of known cosmological parameters---the dimensionless amplitude of the power spectrum of cosmological perturbations $\Delta_\Rcal^2$ and the baryon-to-photon number ratio $\eta_\text{b}$:
\begin{equation}
    \frac {\Omega_\text{DM}}{\Omega_\text{b}} \sim \frac {\Delta_\Rcal^2}{\eta_\text{b}}\,,
\end{equation}
with the only additional requirement being that baryon-antibaryon annihilation be suppressed in the dark sectors.

In our species, the plasma is dense and thermal equilibrium is achieved: it is key that the majority of baryons annihilate with antibaryons, creating the observed small baryon-to-photon number ratio. In the other species, given a large enough $N$, the individual partial densities are extremely low. We show that all annihilation processes are frozen out at all times, which ensures that all the dark baryons survive.

The combination of: couplings bounded by unitarity, constraints on the modulator vacuum expectation value (``VEV'') resulting from the observed cosmological-perturbation amplitude, and baryon asymmetry as given by the observed baryon-to-photon number ratio in our model predicts the correct dark-matter abundance when inflation occurs around the natural cut-off scale.

The paper discusses the scenario in detail: requirements for the inflationary mechanism and the mechanism for generating perturbations are considered in section~\ref{sub:infl}, while the reheating process, generation of the dark matter and the subsequent evolution of the baryonic and dark sectors are laid out in section~\ref{sub:reh}.

\section{Inflation and Perturbations}
\label{sub:infl}
\subsection{Inflationary Background}
We consider a theory with a large number, $N$, of standard model copies, which we label by an index  $j\, = \, 1,2,\ldots N$.  For maximal predictivity and simplicity, we assume that each of the copy sectors is an exact replica of the SM, and all such sectors are related by an exact permutation symmetry $P(N)$. Motivated by the suggested solution of the hierarchy problem \cite{Dvali:2007hz, Dvali:2007wp}, we keep in mind  that the number of species can be as large as $N \, \sim \,  10^{32}$. However, we show that the scenario proposed here also works for a range of values of $N$. Therefore, for generality, we will keep $N$ as a free parameter in most of the below.

It is widely accepted (see textbook discussion e.g.  \cite{Linde:2005ht,Mukhanov:2005sc,Weinberg}) that the history of the early universe starts with a period of inflation  driven by an effective scalar degree of freedom $\Phi$, which has a sufficiently flat potential that allows the field to slow roll. Taking in mind the benefits of inflation  we shall not question this paradigm, and supplement the SM fields by the inflaton \footnote{It would be interesting to consider whether this role could be played by the Higgs field.}. Our $N$-species framework allows us to endow each $j$-th SM copy with its own inflaton field  $\Phi_j$, enumerated by the same index $j$.

At the level of the lagrangian, there is a full permutation symmetry between the SM copies and the associated inflatons. It is clear from our experience of the world that this symmetry must be broken by the cosmological background since we are coupled to a fraction larger than one in $N$ of the matter in the universe. We posit that the seed of this symmetry breaking was prepared by inflation. All we require is that the evolution of the last few e-foldings of inflation be driven by a single inflaton species. Such a situation is much more natural than one may naively think:

The reason is the bound \eqref{bound} applied to the inflationary Hubble parameter, which tells us that, unless the individual inflatons give small ($\ll M_*$) contributions to the Hubble parameter, inflation can be driven at most by a few inflatons simultaneously.  To illustrate this point, consider a simple case---the potential,
\begin{equation}
\label{potential}
    V=\sum_j V(\Phi_j),
\end{equation}
where equality of all the individual potentials arises owing to the permutation symmetry. We assume that each individual sector is capable of driving inflation in the absence of contributions from the others, just as in minimal inflationary scenarios:  for the right choice of inflaton VEV, the would-be Hubble parameters $H_i^2 = V(\Phi_i)/3\Mpl^2$ are not much below $M_*$ and the potentials have appropriate slow-roll parameters.

\begin{figure}[t]
\psfrag{L}[l]{$\lambda_{22}$}
\psfrag{L1}[r]{$\lambda_{22}$}
\psfrag{Q1}[b][l]{$\Phi_{i}$}
\psfrag{Q2}[b][r]{$\Phi_{j}$}
\psfrag{n}[t]{$n$-loops}
\psfrag{F}[b]{$\Phi_k$}
\psfrag{Fd}[t]{$\Phi_k$}
\includegraphics[width=\columnwidth]{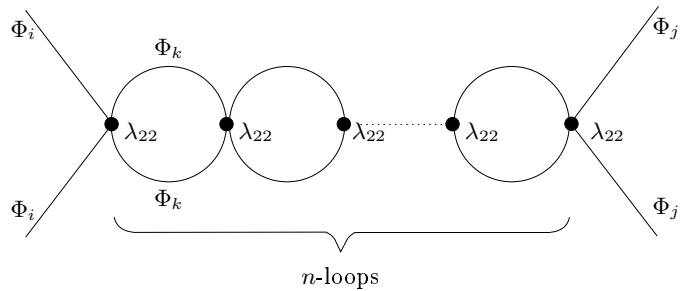}
\caption{Introducing a cross-species coupling to the inflaton potential involves diagrams which couple species through loops. Since the species index on the loops is free, these amplitudes are relatively enhanced by a factor $N$ for every loop, $\mathcal{M} \sim \lambda_{22}(\lambda_{22}N)^n$. Perturbative unitarity of the theory therefore requires that the coupling $\lambda_{22}$ be suppressed by at least a factor $N^{-1}$.}
\label{fig:bubble}
\end{figure}

In such a case, there would exist an inflationary state where inflation is being driven by all $N$ inflatons, giving an effective Hubble parameter of $H^2_\text{eff} = \sum_j V(\Phi_j)/3\Mpl^2 \sim  N  M_*^2$. This is far in excess of the bound~\eqref{bound} and therefore inconsistent. The system should contain a built-in mechanism to prevent such a collective inflationary state: we can add cross-couplings between the species to achieve this, for example,
\begin{align}
\label{potential1}
    V \, =& \,  \sum_j  \, V(\Phi_j)  \, +   \, \lambda_{22}
    \,  \sum_{j_1\neq j_2}\Phi_{j_1}^{2} \Phi_{j_2}^{2} \,  + \, \notag\\
    &+\, {\lambda_{24}  \over M_*^2} \,   \sum_{j_1\neq j_2}\Phi_{j_1}^{2} \Phi_{j_2}^{4} \,  + \, \text{higher terms} \,.
\end{align}
In the case of a cross-species coupling, there are new constraints that arise as a result of the requirement for the perturbative unitarity of the theory. For  example, consider the simplest cross-coupling $\lambda_{22} \Phi_i^2\Phi_j^2$. Having such a coupling allows us to consider annihilation diagrams for a particular species $i$ to a species $j$ mediated by loops of all other species (see Fig.~\ref{fig:bubble}). Since the species index on each loop is free, such diagrams will involve summations over all the species, and therefore will be relatively enhanced with respect to one-specie theories. Diagrams with $n$ consecutive loops are possible and will scale as $\lambda_{22}(\lambda_{22}N)^n$. This indicates that we must constrain $\lambda_{22} \lesssim N^{-1}$ in order to prevent the amplitudes from diverging.

Similar bounds follow from many other considerations including the renormalization of the kinetic terms of the species due to loops of other species.   This scaling is also fully compatible with the typical values of the cross-couplings induced via the exchange of virtual black holes \cite{Dvali:2007wp}: the dominant contribution comes from black holes of mass $M_*$ and   each exchange by a virtual black hole between the different species requires a suppression by $1/N$ in the amplitude.

Let us assume that all the couplings are positive-definite. In that case, despite the required suppression of the cross couplings, putting one of the inflatons on the inflationary slope at some VEV, say $\Phi_k \gtrsim \Mpl$, forces all the other inflatons to remain at zero VEV or, if they had a non-zero VEV before, to occupy zero VEV during the time of order $H^{-1}$. Indeed, the effective mass term for all inflatons with index $j\neq k$ in the background of $\Phi_k$ is given by
\begin{equation}
    \label{mass1}
  M^2_{\Phi_{j\neq k}} \, = \,  \pdbd{^2V(\Phi_j)}{\Phi_j^2} + 2\sum_\alpha \frac{ \lambda_{2,2\alpha}}{M_*^{2\alpha-2}}\Phi_k^{2\alpha}
\end{equation}
This effective mass is at least of order $M_*$, and thus all the other inflaton fields strongly violate the slow-roll condition: the effective friction cannot prevent them from fast roll towards the minimum. At the same time, quantum perturbations for the degrees of freedom with mass $M_{\Phi_{j\neq k}}>H_*$ are not enhanced, leaving no way to produce a VEV for them.

Moreover, non-perturbative black-hole arguments can be used to derive a bound in (quasi) de-Sitter spaces on the masses of particles which have $N$ species \cite{Dvali:2008sy},
\be
\label{boundM}
    m < \frac{\Mpl}{\sqrt{N}}\,.
\ee
Any inflationary vacuum in which this bound is not satisfied must destabilize and cannot support slow-roll. If the coupling $\lambda_{22}$ exceeded $1/N$, the inflationary vacuum with $\Phi_k \gg \Mpl$ would result in the presence of order $N$ species with masses $\gg \Mpl/\sqrt{N}$, violating the bound above.  This argument is fully non-perturbative and is independent of the unitarity considerations, but results in the same constraint on the cross-species couplings as those we found through unitarity argument above.

The system of cross-couplings also eliminates the undesired possibility of inflating collectively with all inflatons at small VEVs, but with $\sum_j \Phi_j^2 \sim \Mpl^2$. Such a situation also provides sufficient inflation for at least some forms of the potential Eq.~\eqref{potential}, but reheats to all the species equally. With the cross-couplings present, the cross-terms add up in such a way that the mass of all the inflatons is high and no inflation can occur.

Overall, despite the symmetry of the potential, the presence of the cross couplings ensures that inflation can only proceed in one of the inflaton species but not several simultaneously.

We can further generalize the cross-coupling terms in Eq.~\eqref{potential1}:
\begin{equation}
V  = \sum_i V(\Phi_i) + \sum_{\alpha_1,...\alpha_n}  {\lambda_{\alpha_1,...\alpha_n}
\over M_*^{\alpha_1 + ... \alpha_n - 4}}  \sum_{j_1,...j_n}\Phi_{j_1}^{\alpha_1}\, ... \, \Phi_{j_n}^{\alpha_n}.
\nonumber
\end{equation}
Here, the first term is the sum over the individual inflaton potentials, while the second term is the sum over the cross couplings, meaning that in each term at least two or more distinct species participate  (that is $n> 1$).  In the presence of individual quantum numbers of the inflatons, the powers $\alpha_k$ are constrained by symmetries.

As before, unitarity arguments constrain these multi-specie cross-couplings to scale with $N$ as  $\lambda_{\alpha_1,...\alpha_n} \, \lesssim \,  N^{1-n}$, i.e.\ couplings between $n$ different species need to be suppressed to at least a factor $N^{1-n}$ since, for each species, a loop can be formed and summed over.

The particular choice of potentials discussed here and the chosen cross-couplings serve only an illustrative purpose, capturing the essence of the mechanism through which the fundamental theory can prevent collective inflation in order to obey the bound on the Hubble parameter. From here on, we shall keep the form of the individual inflaton potentials fully generic. For the purpose of model building, notice that there is no inconsistency in having inflaton VEVs significantly above $M_*$, as long as the value of the Hubble parameter does not exceed the fundamental scale $M_*$ \footnote{The question of whether the inflaton VEV can exceed $\Mpl$ is separate and more profound. The impossibility of this can be proven in certain cases and results from the conflict with black-hole thought experiments that one can perform in de-Sitter space \cite{Dvali:2008sy}.  In the context of this work, we will not worry about this issue, since inflation models with sub-Planckian field VEVs can be constructed.}.

\subsection{Perturbations}

The main benefit of inflation is a natural quantum mechanical generation \cite{Mukhanov:1981xt} of primordial cosmological perturbations which are necessary to seed the inhomogeneities observed in the universe.  For a modern pedagogical treatment of this mechanism and historical details see e.g.
\cite{Linde:2007fr,Mukhanov:2005sc,Weinberg}.

The generated inhomogeneities are usually characterized by the dimensionless power spectrum $\Delta_\Rcal^2(k)$ of curvature perturbations $\Rcal$. In models of inflation involving a single scalar field with a canonical kinetic term, this power-spectrum is given by:
\begin{equation}
\Delta_\Rcal^2(k) = \frac{H_*^2}{8 \pi ^2 \epsilon\Mpl^2},
 \label{ampl}
\end{equation}
where $H_*$ is the Hubble parameter at the time of horizon exit of a particular mode $k$, $\epsilon\equiv-\dot H /H^2$ is a slow-roll parameter,  and we have used the reduced Planck mass $\Mpl = (8 \pi G_N)^{-1/2}$. The function $\Delta_\Rcal^2(k)$ has usually only a slow logarithmic dependence on the scale $k$ which can be approximated in the narrow observable range by a power law $\Delta_\Rcal^2(k)\propto k^{n_S-1}$. One of the spectacular successes of observational cosmology is the measurement by the Wilkinson Microwave Anisotropy Probe (``WMAP'') \cite{Komatsu:2008hk} of the spectral index, $n_S=0.960\pm0.013$, and the value of the power spectrum, $\Delta_\Rcal^2 = (2.445\pm0.096) \times 10^{-9}$, at the scale $k=0.002\ \text{Mpc}^{-1}$. This measurement is in an excellent agreement with predictions of the simplest inflationary models.

For a larger number of inflaton species, the result of Eq.~\eqref{ampl} is not modified \cite{Liddle:1998jc,Dimopoulos:2005ac}, see also \cite[page 506]{Weinberg}. Thus, in chaotic inflation, in the case of many species, where the Hubble constant is bounded by the fundamental scale \eqref{bound}, one obtains
\begin{equation}
\Delta_\Rcal^2(k) = \frac{H_*^2} {8 \pi ^2\epsilon N M_*^2}<\frac{1} {8 \pi ^2\epsilon N }.
\label{ampl2}
\end{equation}
From this inequality, it follows that only models with `moderate' values of $N$ ($N \lesssim  10^{10}$) can avoid extremely small values of the slow-roll parameter $\epsilon$ and potential tensions with the measurement of $n_S$ quoted above. This bound for the number of species in inflation can be obtained in different ways see e.g. \cite{Dvali:2008sy,Huang:2007zt,Ahmad:2008eu,Adshead:2008gk}. In theories with these moderate numbers of species, cosmological perturbations can be generated in the standard manner and no alternative mechanisms are required. However, for larger values of $N$,  inflation is forced to proceed at a lower scale and the amplitude of density perturbations generated in standard chaotic inflation becomes unacceptably small.

The issue of generating density perturbations in the presence of a large number of species is studied in \cite{Sawicki:2009}. A number of successful scenarios exist, but the mechanism most suitable for our purposes is modulated reheating \cite{Dvali:2003em,Dvali:2003ar}, which enables the generation of sufficient, red-tilted cosmological perturbations for the small values of $H_*$ required by the bound Eq.~\eqref{bound}.
In this mechanism, one introduces a new scalar field $\chi$, the \emph{modulator}, which controls the decay rate, $\Gamma$, of the inflaton field into SM particles. This modulator $\chi$ is assumed to be light, $M_\chi\ll H_*$, so that substantial quantum fluctuations $\delta\chi \sim H_*$ are generated during inflation. These fluctuations become frozen (classical) as they leave the horizon. Moreover, it is assumed that $\chi$ does not contribute significant energy density during inflation and does not affect inflationary dynamics:
\be\label{chimass}
    H_*^2 \gg \frac{M_\chi^2\chi^2}{\Mpl^2}.
\ee

As usual, the inflationary epoch is followed by the oscillatory dust-like stage. During this stage the Hubble constant decreases quickly, $\epsilon>1$, and eventually becomes less than the decay rate $\Gamma$. Around that time reheating occurs: inflaton decay becomes efficient enough to populate the universe with SM particles. Without loss of generality, we assume that during reheating the inflaton(s) oscillate about $\Phi_j = 0$. The local VEV $\left\langle\chi\right\rangle$ controls the decay rate of the inflatons, $\Gamma$, and therefore controls the position of the reheating surface. Perturbations in the field value of the modulator lead to perturbations in the reheating surface. Assuming that the universe was dust-dominated prior to reheating, radiation-dominated afterwards and that reheating occurs instantaneously at $H\simeq\Gamma$, one obtains \cite{Dvali:2003em,Dvali:2003ar} for the curvature perturbation in the radiation fluid
\be\label{zeta}
    \Rcal = -\frac{1}{6}\frac{\delta\Gamma}{\Gamma} = -\frac{1}{3}\frac{\delta\chi}{\left\langle\chi\right\rangle}\,,
\ee
where the final equality results from the fact that the couplings which provide the dominant decay channels are linear in the modulator field. We have also assumed that the amplitude of perturbations in the inflatons is negligible compared to that imparted to the modulator.

We require that the modulator $\chi$ be common for all species. Thus all the perturbations in the decay products of the inflaton are generated by a single source: perturbations in the modulator field. This ensures that, even if the decay products are not thermally coupled after reheating, their curvature perturbations will be identical and will not contain an isocurvature mode. The relative contributions to the energy density in different particle types will be the same everywhere, only the time of reheating will have changed.

The modulation of the reheating process can be guaranteed by invoking a symmetry which ensures that all inflatons couple to the light SM fields through the modulator $\chi$. The simplest possibility is to require a $\mathbb{Z}_2$ symmetry that we shall refer to as inflaton parity. We shall assume that this symmetry is common for all the copies. Under this symmetry, all the inflatons, as well as the modulator, change sign,
\be\label{parity}
 \Phi_j \, \rightarrow - \Phi_j \qquad\qquad \chi \, \rightarrow \, -\chi \,,
\ee
while all the usual SM fields and their replicas are left invariant. This fact ensures that the decay rate of all the inflatons is modulated by the fluctuations of the expectation value of $\chi$.

The dimensionless curvature power spectrum resulting from this reheating mechanism is given by
\be
    \Delta_\Rcal^2 = \frac{H_*^2}{(6\pi)^2\etion{\chi}^2}.
\ee
For the correct amplitude of the curvature perturbations, the vacuum expectation value for the modulator during inflation is then required to be
 \begin{equation}
\label{chivev}
\left\langle\chi\right\rangle \, \sim \, \frac{H_*}{\Delta_\Rcal} \, \sim \, 10^5\,  H_* \,.
\end{equation}

\section{Reheating and Freeze Out}
\label{sub:reh}
\subsection{Inflaton Decay}

The lowest-order invariant couplings between the inflatons and the SM species
(call them $Q_j$)  can be parameterized in the following way
\begin{equation}
\label{couplings}
\Phi_i \,{\chi  \over M_*} \,   \left ( g\, Q_iQ_i \, + \,\tilde{g} \,Q_jQ_j \right) \, ,
\end{equation}
where the two coupling constants $g$ and $\tilde{g}$ are responsible for the diagonal and non-diagonal decays respectively.
Notice that, owing to the underlying structure of the theory, these effective coupling constants may depend on some VEVs. For example, if $Q$ are quarks and the inflaton is a SM gauge singlet, the coupling constants may depend either on the Higgs VEV or on powers of the four-momentum.  The convenience of the above parameterization is that such details are unimportant: we express everything in terms of the cutoff and the effective dimensionless couplings which must satisfy the unitarity constraints, irrespective of their substructure.

In our analysis, we shall assume that the values of these coupling constants are the most natural (the largest possible) compatible with the requirement of unitarity at the energies below the cutoff scale. We discuss these unitarity bounds below.

\begin{figure}[t]
\psfrag{A}[t]{$(a)$}
\psfrag{B}[t]{$(b)$}
\psfrag{Q1}[b][l]{$Q_{i}$}
\psfrag{Q2}[b][l]{$Q_{i}$}
\psfrag{X}[b]{$\chi$}
\psfrag{Phi}[b]{$\Phi_k$}
\psfrag{GL1}[t]{\large$\frac{g}{M_{*}}$}
\psfrag{GR1}[t]{\large$\frac{g}{M_{*}}$}
\psfrag{cl}[t]{\large$\frac{\tilde{g}\left\langle\chi\right\rangle}{M_{*}}$}
\psfrag{cr}[t]{\large$\frac{\tilde{g}\left\langle\chi\right\rangle}{M_{*}}$}
\psfrag{Phi_in}[b][l]{$\Phi_{i}$}
\psfrag{Phi_out}[b][r]{$\Phi_{j}$}
\psfrag{Q3}[b][r]{$Q_{j}$}
\psfrag{Q4}[b][r]{$Q_{j}$}
\includegraphics[width=3.2in]{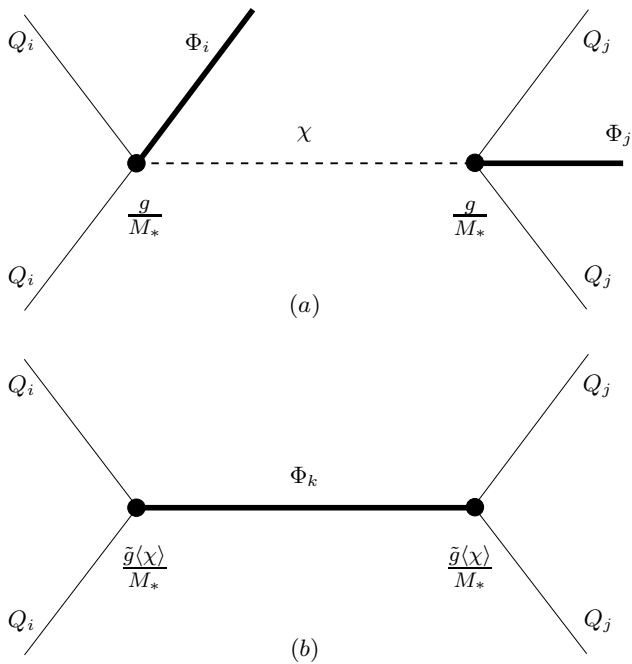}
\caption{Processes which lead to strongest constraints on the coupling constants $g$ and $\tilde{g}$ arising as a result of interaction terms Eq.~\eqref{couplings}. Bounds are governed by the requirement for perturbative unitarity. (a) Exchange of a virtual $\chi$. (b) Exchange of a virtual inflaton $\Phi_k$; total amplitude involves a sum over all inflatons $k$.}
\label{dia}
\end{figure}

As was already explained in the previous section, the general $N$-scaling of the inter-species couplings is such that an operator involving  $n$ different types of species must  be suppressed by  $1/N^{n-1}$. The most stringent unitarity bound on $g$ comes from this consideration.  Even though this operator is diagonal in the species index, it can generate inter-species interactions via  the exchange of a virtual $\chi$. Such a process is shown in  Fig.~\ref{dia}(a). At characteristic energies $E<M_\chi$, we can integrate the modulator out to obtain an effective six-point interaction between the two types of species,  with the vertex $g^2/M_*^2M_{\chi}^2$.   Assuming that the mass parameters are of roughly the same order,  we get the following constraint,
\begin{equation}
\label{bound1}
    g  \lesssim  N^{-1/2}\,.
\end{equation}

The bound on the non-diagonal coupling, $\tilde{g}$, is at least $1/N$, since it involves species with two different labels.  However, a more stringent constraint can be obtained since the modulated-reheating mechanism requires the presence of a background field VEV, $\etion{\chi}$. On such a background, we can reduce the $\tilde{g}$ vertex of  Eq.~\eqref{couplings} to a three-point interaction. The unitarity bound then arises from the diagram involving the exchange of virtual-inflaton species, shown in Fig.~\ref{dia}(b), with the effective coupling $N\tilde{g}^2\etion{\chi}^2/M_*^2M_\Phi^2$, where the factor $N$ results from the summation over all the species of virtual inflaton. The constraint on the coupling then becomes
\be
    \tilde{g}\, \lesssim \, \frac{M_*}  {\etion{\chi}} \, N^{-1}\,.
\ee
Using \eqref{chivev}, this constraint can be recast as
\begin{equation}
\label{bound2}
\tilde{g}\, \lesssim \, \Delta_\Rcal \, N^{-1}\left(\frac{M_*}{H_*}\right)  \sim 10^{-5} N^{-1} \,  \,.
\end{equation}

One can also constrain the maximal inflaton VEV during inflation. Namely, unitarity arguments are most restrictive for the rate of the process $Q_1+Q_1\xrightarrow{\chi} Q_j+Q_j$ to any $j$, occurring in the presence of $\etion{\Phi_1}$. For the characteristic energies $E$,  the rate is bounded in such a way as to satisfy $\Gamma(E)<E$. Demanding that unitarity hold up to the cutoff scale and using the constraints on $g$ and $\tilde g$, bound \eqref{bound1} and bound \eqref{bound2} respectively, one obtains
\be
\frac{\etion{\Phi_1}}{\Mpl} \lesssim \sqrt{\frac{\etion{\chi}}{M_*}}\sim 100.
\ee
Note that this bound is not restrictive at all.

As described in section~\ref{sub:infl}, the decay products after reheating all come from a single inflaton species, $\Phi_1$. The coupling $g$ is relatively large, Eq.~\eqref{bound1}, and controls the diagonal decay into matter species $Q_1$. However, particles in the other copies are also produced by virtue of the much smaller off-diagonal coupling $\tilde{g}$, Eq.~\eqref{bound2}. The individual energy densities in all the other sectors are negligible relative to our SM sector; however, the integrated energy density is significant. The ratio of the energy densities between the SM and any other given $j$-th copy is set by the ratio of the inflaton decay rates into the two sectors
\begin{equation}
\label{rates}
    {\rho_j \over \rho_1} \, = \, {\Gamma_j \over \Gamma_1} \, = \, {\tilde{g}^2 \over g^{2}}\, \sim \, \frac{M_*^2}{N\etion{\chi}^2}\, \sim \, \,\left(\frac{M_*}{H_*}\right)^2  \Delta_\Rcal^2 \, N^{-1} \,.
\end{equation}
where we have used Eq.~\eqref{chivev}. Or, equivalently, the total energy density in the dark sector right after the reheating is
\be\label{darkdens}
    \sum_{j\neq 1}\rho_j \sim \rho_1 \left(\frac{M_*}{H_*}\right)^2 \, \Delta_\Rcal^2 \sim 10^{-10} \rho_1,
\ee
and is strongly suppressed compared to the energy density in our sector. Note that this result does not depend on the number of species $N$. Since interactions between the sectors are suppressed, the other species are never in thermal equilibrium with our sector. This means that they are not included in the naive tally of energy density in radiation and therefore contribute an energy density that is unaccounted for during Big-Bang nucleosynthesis. The smallness of Eq.~\eqref{darkdens} guarantees that the nucleosynthesis bound on the amount of extra radiation in the Universe \cite{Barger:2003zg} is satisfied.

\subsection{Reheating in Our Sector}

Let us now estimate whether we can reheat the universe to a sufficiently high temperatures $T_\text{RH}$ to synthesize the light elements in successful Big-Bang Nucleosynthesis (``BBN'').
The reheating takes place once the decay rate of the inflaton $\Gamma_{\Phi_1}  \, \sim \, H$. The branching ratios to species other than $j=1$ are subleading, so we can use the approximation that the total decay rate is equal to the decay rate to our species,
\be
\label{Gamma}
\Gamma_{\Phi_1} \sim \Gamma_1 =\frac{g^2\etion{\chi}^2}{8\pi M^2_*} M_{\Phi _1},
\ee
is the decay rate of the inflaton: $\Phi_1\xrightarrow{} Q_1+Q_1$.
Therefore, right before reheating, the energy density stored in the inflaton $\Phi_1$ is
 \begin{equation}
\label{infenergy}
\rho_{\Phi}\,  \sim \, \Gamma^2_{1} \Mpl^2 \, \sim   \frac{1}{N^2\Delta^4_\mathcal{R}}\left(\frac{H_*}{M_*}\right)^4 \Mpl^2 M_{\Phi_1}^2\,,
\end{equation}
where we have used Eq.~\eqref{bound1} and Eq.~\eqref{chivev}.
Then, within  time $\sim \Gamma_1^{-1}$, the energy of inflaton oscillations gets converted into the energy of the species.   During the decay, $Q_i$ are created with the energy of order $E \, \sim \, M_{\Phi_1}$. This is true for all the species.  The number density in our sector, $n_1$, is large and allows for rapid thermalization.

Let us estimate the thermalization rate. This is the same as the interaction rate of SM particles:
\begin{equation}
\label{thermalrate}
\Gamma_\text{T} \, \sim \, n_1 \sigma \, \sim {\rho_\Phi \over M_{\Phi_1}} {\alpha^2 \over M_{\Phi_1}^2} \,
\sim \frac{\alpha^2}{N^2\Delta^4_\mathcal{R}}  \left(\frac{H_*}{M_*}\right)^4 {\Mpl^2 \over M_{\Phi_1}} \,.
\end{equation}
Since, from \eqref{infenergy} the Hubble parameter at reheating is,
\begin{equation}
\label{hubblereheat}
H_\text{RH} \, \sim \Gamma_1\sim\, M_{\Phi_1}\left(\frac{H_*}{M_*}\right)^2 \frac{1}{N\Delta^2_\mathcal{R}} \, ,
\end{equation}
we obtain the result that the number of scatterings per particle $Q_1$ during one Hubble time is
\begin{equation}
\label{condition1}
{\Gamma_\text{T} \over H_\text{RH}} \, \sim  \,  \frac{\alpha^2}{ N\Delta_\mathcal{R}^2}  \left(\frac{H_*}{M_*}\right)^2 \left (\frac{\Mpl}{M_{\Phi_1}} \right)^2.
\end{equation}
Using \eqref{bound} this expression above can be rewritten in the form
 \begin{equation}
\label{condition2}
{\Gamma_\text{T} \over H_\text{RH}} \, \sim \, \frac{\alpha^2}{\Delta_\mathcal{R}^2}  \left ( {H_* \over M_{\Phi_1}} \right )^2 .
\end{equation}
In our inflationary scenario, $H_* > M_{\Phi_1}$, hence
\be
{\Gamma_\text{T} \over H_\text{RH}} \, \gtrsim \, \frac{\alpha^2}{\Delta_\mathcal{R}^2} \gg 1,
\ee
and the thermalization happens well within a Hubble time.  Therefore, the reheating temperature can be estimated using
\begin{align}
\label{reheatest}
H_\text{RH}\sim \frac{T^2_\text{RH}}{ \Mpl} \sim \Gamma_1 \sim \frac{g^2\etion{\chi}^2M_{\Phi _1}}{M^2_*},
\end{align}
which combined with Eq.~\eqref{chivev}  yields
\begin{align}
\label{TRH}
T_\text{RH} \sim \frac{g}{\Delta_\Rcal} \left(\frac{H_*}{M_*}\right) \sqrt{\Mpl M_{\Phi _1}}.
\end{align}
Further, we apply the constraint Eq.~\eqref{bound1} along with the bounds on the Hubble parameter (1), $H_*<M_*$, and on the inflaton mass Eq.~\eqref{boundM}, $M_{\Phi _1}<M_*$, and obtain the following upper bound for the reheating temperature
\begin{align}
T_\text{RH} \lesssim \frac{\Mpl}{\Delta_\Rcal}N^{-3/4} \sim 10^{5} \Mpl N^{-3/4}.
\label{Temperature}
\end{align}
For $N \sim 10^{32}$ the reheating temperature is around or less than 1 GeV and it is only higher for smaller $N$.
In particular,  for $N \sim 10^{20}$ we have $T_\text{RH} \sim M_* \sim M_{\Phi _1} \sim 10^9$ GeV. It is worth mentioning that $T_\text{RH} \sim 10^5M_*N^{-1/4}$, therefore for all $N\gtrsim 10^{20}$ the reheating temperature is much less than the assumed mass of the inflaton $M_{\Phi _1} \lesssim M_*$.
The number of species $N$ clearly cannot exceed $10^{32}$ by more than a few orders of magnitude, since otherwise the strong-gravity effects would have already been seen by accelerators. Hence, in our scenario $T_\text{RH}\gtrsim1$ GeV for all interesting $N$, which ensures that BBN proceeds unaltered.

Thus, shortly after reheating, the universe is left with our sector at temperature $T_\text{RH}$, while the other sectors---with much lower particle densities---remain unthermalized and consist of a collection of particles with energies $\sim  M_{\Phi_1}$.

The reason for the final domination of the dark-matter sector over the baryons lies in the difference in their behavior resulting from the huge hierarchy in the partial energy densities of the individual sectors.

The thermal history for our SM sector is unaltered: The high density of SM baryons and anti-baryons ensures that they remain thermally coupled, converting the baryons and antibaryons into photons extremely efficiently. Only as a result of some primordial baryon/anti-baryon asymmetry does a significant number of baryons survive. Nonetheless, these represent a tiny fraction of the original number, which is known to be
\be\label{eta}
    \eta_\text{b} = \frac{n_\text{b}}{n_\gamma} = 2.7\cdot  10^{-8}\, \Omega_\text{b} h^2 = 6.2 \cdot 10^{-10}\,.
\ee
This number is also known to not have been significantly different as early as at the time of nucleosynthesis \cite{Weinberg}. Therefore, for every baryon in our species that exists today, approximately $\eta_\text{b}^{-1}$ baryons in our copy of the SM were in existence after reheating.

\subsection{Dark Sector Freeze-Out}

For the baryons not in our species, the story is very different: given a large enough $N$, their partial densities following reheating are extremely low. In our model, we require that they be so rare that all annihilation processes are always frozen out and the other species never reach thermal equilibrium. We define freeze-out as the time at which the annihilation rate for the $j$-th copy, $\Gamma^\text{A}_j$, falls below the Hubble parameter.

Let us consider the annihilation of hidden protons and antiprotons in the species $j$ at the time when the the temperature of the particles in our sector is $T$. The number of annihilations per particle during one Hubble time is given by
\begin{equation}
\label{annihil}
  \frac{\Gamma^\text{A}_j} {H} \simeq \frac{n_j \sigma v}{H}\,,
\end{equation}
where $n_j$ is the baryon number density of the dark species in sector $j$, $\sigma$ is the annihilation cross-section and $v$ is the typical speed of the particle. The density $n_j$ is conserved in a comoving volume, provided that annihilations be suppressed. We can, therefore, express $n_j$ in terms of the density of our baryons as
\begin{equation}
  n_j(T) = n_1 \frac{\rho_j}{\rho_1} \left(\frac{T}{T_\text{RH}}\right)^3  \,,
\end{equation}
where $n_1$, $\rho_j$  and $\rho_1$ are evaluated before thermalization occurs. In particular, the number density in our species can be estimated using Eq.~\eqref{infenergy} and Eq.~\eqref{reheatest}
\begin{equation}
  n_1\simeq \frac{\rho_{\Phi}}{M_{\Phi_1}} \sim \left(\frac{T_\text{RH}}{M_{\Phi_1}}\right) T^3_\text{RH}.
 \end{equation}
Thus, using Eq.~\eqref{rates}, we obtain the $j$ baryon number density  as the function of the temperature in our thermalized sector
\be
\label{jbaryon}
n_j(T)=\left(\frac{T_\text{RH}}{M_{\Phi_1}}\right) \left(\frac{M_*}{H_*}\right)^2 { \Delta_\Rcal^2 \over N} \ T^3.
\ee

There are three primary channels through which mass stored in protons can be reduced: (a) annihilation through a virtual photon to charged particles (electrons, positrons, pions, etc): $j\bar{j}\rightarrow \text{e}^-\text{e}^+$, (b) annihilation to two photons: $j\bar{j}\rightarrow 2\gamma$, (c) formation of a bound state through a radiative-capture process followed by annihilation: $j\bar{j}\rightarrow bound$. There are also three regimes which we will need to consider for the cross-sections: relativistic, non-relativistic and ultra non-relativistic or Sommerfeld enhanced \cite{Sommerfeld}. As we have already mentioned, at preheating, the characteristic energy of a proton in the $j$-th copy of SM is $E(T_{\text{RH}})\sim M_{\Phi_1}\gg m_{\text{p}}$, where $m_\text{p}$ is the mass of the proton. Thus, copies of the protons are initially highly relativistic.

During the expansion of the universe, the momentum of the protons in the $j$-th copy of the SM redshifts, $\vec{p} \propto a^{-1}$. This implies that, while they are relativistic, the characteristic energy redshifts as $E(T)\simeq M_{\Phi_1}T/T_\text{RH}$. Eventually the kinetic energy becomes comparable to the rest mass of the proton and it becomes non-relativistic; this occurs when the temperature in our sector is
\be
\label{critical}
T_\text{cr} \simeq \left(\frac{m_\text{p}}{M_{\Phi_1}} \right) T_\text{RH} \sim \frac{m_\text{p}}{\Delta_\mathcal{R} N^{1/4}} \left(\frac{H_*}{M_*}\right)\sqrt{\frac{M_*}{M_{\Phi_1}}}\,.
\ee
At temperatures below $T_\text{cr}$, the momentum continues to redshift which leads to a redshifting of velocity, $v \sim T/T_\text{cr}$. Finally, when the speed of the dark-sector protons falls beneath the fine-structure constant, $v \lesssim \alpha$, Sommerfeld enhancement becomes important for the cross-sections \cite{Sommerfeld}. The temperature in our sector at this transition is
\be
\label{sommerfeld}
T_\text{Som} \simeq \alpha T_\text{cr}.
\ee

We will now consider the freeze-out condition for each of the three annihilation channels in turn. Using $H\sim T^2/\Mpl$ and Eq.~\eqref{jbaryon} in Eq.~\eqref{annihil} yields
\be\label{GHs}
  \frac{\Gamma^\text{A}_j} {H} \simeq \sigma v \frac{\Delta^2_\mathcal{R}}{N} \left(\frac {T_\text{RH}}{M_{\Phi_1}}\right) \left( \frac{M_*}{H_*} \right)^2 \Mpl T.
\ee
We have presented the parametric forms of the cross-sections $\sigma v$ in Table~\ref{sigmav}. Using these in Eq.~\eqref{GHs} allows us to investigate the evolution of the freeze-out condition for each of the channels.

\begin{table}[ht]
\begin{tabular*}{\columnwidth}{@{\extracolsep{\fill}} l|ccc}
\hline\hline
\textbf{Regime} & \textbf{Charged} & \textbf{Photons} & \textbf{Capture}\\
\hline
Relativistic & \Large${\tabfrac{\alpha^2}{E^2}}$ & \Large${\frac{\alpha^2}{m_\text{p}E}}\text{{\normalsize ln}} \left(\frac{2E}{m_\text{p}} \right)$ & --- \\
Non-Relativistic & \Large$\frac{\alpha^2}{m_\text{p}^2}$ & \Large$\tabfrac{\alpha^2}{m_\text{p}^2}$ & \Large$\frac{\alpha^2}{m_\text{p}^2} \left( \frac{\alpha}{v} \right)^{\normalsize 4}$\\
Sommerfeld & \Large$\tabfrac{\alpha^2}{m_\text{p}^2} \left(\frac{\alpha}{v} \right)$ & \Large$\frac{\alpha^2}{m_\text{p}^2}\left(\frac{\alpha}{v} \right)$ & \Large$\frac{\alpha^2}{m_\text{p}^2} \left(\frac{\alpha}{v} \right)$\\
\hline\hline
\end{tabular*}
\caption{Parametric forms of the cross-sections, $\sigma v$, for three potential annihilation channels for protons and antiprotons: (a) annihilation to light charged particles through virtual photon \cite{Itzykson}; (b) annihilation to two photons \cite{Itzykson}; and (c) formation of a bound state through radiative capture \cite{Akhiezer}. We present three limits of the expressions: highly relativistic, valid for temperatures in our sector $T_\text{cr} < T < T_\text{RH}$, non-relativistic, valid for $T_\text{Som} < T < T_\text{cr}$ and Sommerfeld enhanced \cite{Sommerfeld}, valid for $T<T_\text{Som}$.\label{sigmav}}
\end{table}

\paragraph{Annihilation to charged particles $j\bar{j}\rightarrow \text{e}\text{e}^+$.}
In the relativistic regime, $T> T_\text{cr}$, we can estimate $\sigma \sim \alpha^2 / E^2$, by redshifting the energy we obtain $\sigma v \sim \alpha^2 T^2_\text{RH}/\left(M_{\Phi_1}T\right)^2$. Substituting this into Eq.~\eqref{GHs} yields
 \be
 \label{G/H}
  \frac{\Gamma_{(j\bar{j}\rightarrow \text{e}\text{e}^+)}} {H}\sim \alpha^2 { \Delta_\Rcal^2 \over N}  \left(\frac{T_\text{RH}}{M_{\Phi_1}}\right)^3 \left(\frac{M_*}{H_*}\right)^2 \left(\frac{\Mpl}{T}\right).
\ee
Here it is important to note that as a result of the redshifting of the relativistic $E(T)$ the above function grows as the universe cools down. When the temperature in our sector reaches $T_\text{cr}$ the dark baryons enter the non-relativistic regime. There, according to Table~\ref{sigmav}, $\sigma v=const $ and consequently $\Gamma_{(j\bar{j}\rightarrow \text{e}\text{e}^+)} / H$ decreases as $T$. Finally, after the universe cools to  $T_{\text{Som}}$, the Sommerfeld-enhanced regime starts, and the amount of annihilations per particle per cosmic time, $\Gamma_{(j\bar{j}\rightarrow \text{e}\text{e}^+)} / H$ approaches a constant owing to an additional power of $v$ in the denominator of the cross-section. Thus, the maximum of $\Gamma_{(j\bar{j}\rightarrow \text{e}\text{e}^+)} / H$ is reached when the dark baryons are on the verge of becoming non-relativistic, at $T=T_\text{cr}$. If this annihilation channel for the dark baryons is frozen out at this maximum, then this annihilation reaction is frozen out at all times.

It is convenient to rewrite the expression \eqref{G/H} using the formula \eqref{TRH} and Eq.~\eqref{bound1}
\be
\frac{\Gamma_{(j\bar{j}\rightarrow \text{e}\text{e}^+)}} {H}\sim  \frac{\alpha^2}{N^2}\left(\frac{\Mpl}{M_{\Phi_1}}\right)^2 \frac{T_\text{RH}}{T}.
\ee
Substituting the critical temperature in the freeze-out condition yields a bound on the lowest number of species necessary to ensure that annihilations to light leptons in the dark sector be frozen out,
\begin{align}
  N &> \left(\alpha^2 \frac{\Mpl}{m_\text{p}} \frac{M_*}{M_\Phi}\right)^{2/3} \sim 10^9\,.
\end{align}

\paragraph{Annihilation to two photons $j\bar{j}\rightarrow 2\gamma$.}
Analogously to the previous case, for the relativistic regime, we use the cross-section from Table~\ref{sigmav} and the redshift law for the energy. allowing us to write Eq.~\eqref{GHs} as 
    \begin{equation*}
        \frac{\Gamma_{(j\bar{j}\rightarrow 2\gamma)}} {H}\sim \frac{\alpha^2\Delta_\mathcal{R}^2}{N} \frac{\Mpl}{m_\text{p}} \left( \frac{T_\text{RH}}{M_{\Phi_1}}\right)^2 \left( \frac{M_*}{H_*}\right)^2 \ln \left(\frac {2M_{\Phi_1}T}{m_\text{p} T_\text{RH}}\right)\,.
    \end{equation*}
Thus in the relativistic regime $\Gamma_{(j\bar{j}\rightarrow 2\gamma)}/H$ decreases logarithmically as the universe cools. However, the non-relativistic regime is identical to that of  process (a), i.e.\ $\Gamma_{(j\bar{j}\rightarrow 2\gamma)}/H$ first decreases as $T$ and then becomes constant, once Sommerfeld enhancement is active. Therefore, $\Gamma_{(j\bar{j}\rightarrow 2\gamma)}/H$ is maximal at $T_{\text{RH}}$ and consequently $N$ will be most strongly constrained at reheating where
\begin{equation*}
        \frac {\Gamma_{(j\bar{j}\rightarrow 2\gamma)}} {H_\text{RH}} \sim \frac{\alpha^2}{N^{3/2}} \left(\frac {M_*}{M_{\Phi_1}}\right) \left( \frac{\Mpl}{m_\text{p}} \right) \ln \left( \frac{2M_{\Phi_1}}{M_*} \frac{\Mpl}{m_\text{p} \sqrt{N}} \right) \,.
\end{equation*}
Requiring that this process be frozen out at reheating, constrains the number of species to be
\begin{equation}
    N > 10^{11} \,.
\end{equation}

\paragraph{Formation of bound state.}

Once the protons are non-relativistic, it is possible for them to be captured into a bound state. For the purpose of this estimate we have used the asymptotics of the cross-sections obtained in \cite{Akhiezer}. Once the bound state is formed, the decay rate of the proton-antiproton pair is extremely high, $\sim m_\text{p}\alpha^{5}\sim 10^{10}$~GeV \cite{Itzykson}.

    It is clear from the cross-sections presented in Table~\ref{sigmav} that, for $v>\alpha$, the cross-section for capture is highly suppressed compared to the annihilation channels (a) and (b) and therefore the formation of the bound states does not occur provided those channels be frozen out. Once the universe cools further and reaches $v < \alpha$, $\Gamma_{(j\bar{j}\rightarrow bound)} /H$ becomes constant of the same order as that for the other channels. Therefore there are no new bounds on $N$ arising from this channel.

 \paragraph{Annihilation in halos.}

Another important factor to consider is the annihilation of dark protons in galactic halos. Once the halos are formed as a result of non-linear collapse, the density inside them is much higher than the cosmological energy density. This makes it possible for baryons to interact much more efficiently. At $T<T_\text{Som}$, the cross-sections for all the channels discussed above are parametrically equal (see Table \ref{sigmav}), therefore all these processes will place the same constraint on the number of species.

    We are going to estimate an upper bound for the rate of the annihilation processes. A halo can be  defined as a region with average density $\rho_\text{halo} > 200\bar\rho_\text{m}$, where $\bar\rho_\text{m}$ is the background matter density \cite{Cooray:2002dia}. We will assume that from the redshift at which the halos have first formed, $z_\text{form}$, until today, all the mass of the universe is stored in halos. In such a circumstance, we can write down a redshift-dependent annihilation rate
    \begin{equation}
        \Gamma_{\text{halo}}(z) = n_j\sigma v \sim \frac{\rho_\text{halo}(z)\alpha^3}{Nm_\text{p}^3 v(z)}\,.
    \end{equation}
    For the purpose of this upper bound, we will also assume that, throughout the period under consideration, the velocity of protons does not increase as a result of virialization, but remains constant at the value it had before the dark protons fell into the halos, $v\sim\alpha T(z_\text{form})/T_\text{Som}$. With these assumptions, using Eq.~\eqref{sommerfeld} we obtain the result that
    \begin{equation}\label{halogamma}
        \Gamma_{\text{halo}}(z) \sim \frac{200\rho_0(1+z)^3\alpha^3}{m_\text{p}^2T(z_\text{form}) \Delta_\mathcal{R} N^{5/4}}\,.
    \end{equation}
    Since $\Gamma_{\text{halo}}(z)$ represents the rate per particle of interactions at redshift $z$, we can ensure that the annihilation in halos remain insignificant by requiring that the time integral of Eq.~\eqref{halogamma} over the period of the existence of the halos be small, say
    \begin{equation}
    \int^{t(z=0)}_{t(z=z_\text{form})} \Gamma_{\text{halo}}(z(t)) \ds t < 0.01\,.
    \end{equation}
    Using the conservative value of $z_\text{form} \sim 10$, the Hubble parameter today $H_0\sim10^{-34}\,  \text{eV}$  and $T_0\sim 10^{-4}\, \text{eV}$ we obtain the bound 
    \begin{equation}
        N > 10^{11}\,.
    \end{equation}

Putting all of the above results together, we find that the strongest bound on $N$  comes from the annihilation to two photons at the time of reheating. Hence our scenario works for the number of species in the window:
\begin{align}
\label{window}
10^{11}\lesssim N \lesssim 10^{32}.
\end{align}
Nonetheless, provided that some additional mechanism which suppresses the interactions between the dark baryons existed, the annihilations could be frozen out for much smaller $N$, still giving the appropriate dark-matter abundance.

Here, a cautionary note is necessary. In the analysis above we have naively assumed that the inflaton decays practically directly to a small number of dark (anti)protons. This is not unreasonable for $M_{\Phi_1}\sim M_* \sim 1$~TeV or, equivalently, $N\sim 10^{32}$. However, for a smaller number of species $N$, the fundamental scale is far above TeV and new physics  is required once again to solve the hierarchy problem. This new physics could make our assumptions invalid. Another potential deviation from our setup could be caused by the fact that decays of ultra-massive inflatons may result in the production of a very large number of protons per initial inflaton particle. Each dark proton would then carry much less energy than the inflaton mass, contrary to the assumptions that we have made. This would enhance the annihilation rates and therefore make the lower bound on $N$ higher. In particular, if the energy  per proton after inflaton decay is $M_{\Phi_1} / \mu$, then the expressions for the number of annihilations  per cosmic time, e.g. the right-hand side of Eq.~\eqref{G/H}, are enhanced by $\mu^2$. For example for $N=10^{32}$ and $M_{\Phi_1}\sim M_* $, we have $\Gamma_{(j\bar{j}\rightarrow 2\gamma)}/H (T_{\text{RH}})\sim 10^{-33} \mu^2$ while $\mu$ cannot exceed $1000$. Thus even a very high multiplicity would not change our conclusions and can at most affect the lower bound on $N$.

It may seem surprising that dark matter could be baryonic: in the presence of large densities of baryons, the interaction rates are large and they are able to cool and fall into the center of halos to form stars. However, given a large enough number of copies of the SM, $N$, each species becomes so dilute that the probability for any baryon to meet another from its own species becomes extremely small. Since the nucleon density today is approximately $n_\text{b}^0 = 3\cdot 10^{-7}$~cm$^{-3}$, with $N=10^{32}$ we would have a partial density in any other species of approximately $n_j^0 \sim 10^{-38}$~cm$^{-3}$. Unitarity bounds cross-specie couplings to be extremely small and the baryons become nearly collisionless, interacting effectively only through gravity.

The limits on the mass for such dark-matter candidates are weak: constraints arising from the Lyman-$\alpha$ forest limit the mass of a sterile neutrino to be $m_\nu \gtrsim 10$~keV \cite{Seljak:2006qw,Boyarsky:2008xj}. In our proposal, the majority of the mass of the universe is in protons and anti-protons in other species, with their mass $m_\text{p}\sim 1$~GeV easily satisfying the current constraints.

Finally, we can calculate the contribution of such dark matter to the total energy budget of the Universe today. Taking the result Eq.~\eqref{darkdens} for the ratio of the density in our sector to the dark-matter density and adjusting it for the fact that only a fraction given in Eq.~\eqref{eta} of baryons in our sector will fail to annihilate to radiation, we obtain,
\begin{equation}
\label{dm}
\frac{\Omega_\text{DM}}{\Omega_\text{b}} \sim \eta_\text{b}^{-1} \frac{M_*^2}{\etion{\chi}^2} \sim \frac{\Delta_\Rcal^2}{\eta_\text{b}} \left(\frac{M_*}{H_*}\right)^2 \gtrsim 4
\end{equation}
This is rather remarkable: taking the inflationary Hubble parameter and inflaton mass to their most natural value, around the cutoff, together with the saturation of the unitarity constrains on coupling constants gives the correct dark-mater abundance.

\section{Conclusion}
In this paper we have shown that, in models with a large number of species, we can naturally reproduce the observed dark-matter abundance. The parameters of such models are constrained by the requirement for perturbative unitarity: the effective coupling between species must be suppressed in such a way that amplitudes with loops of species do not diverge.

We have introduced a permutation symmetry, which results in our Universe's containing $N$ copies of the standard model. Inflation occurs through the inflaton belonging to our copy, while perturbations and reheating are driven by a modulator field common to all species. This field mediates couplings between the inflaton and both our sector and the other species. The observed amplitude of cosmological density perturbations determines the value of the VEV of the modulator. Combining this with the above-mentioned unitarity constraints produces a natural hierarchy between the inflaton couplings to our sector and to the other species. The amount of matter produced in each species is determined solely by these considerations.

In our species, the density is large and therefore baryon-antibaryon annihilation proceeds unimpeded. Provided that the number of species be sufficiently large, for example appropriate to account for the hierarchy problem, the annihilation in the other species is always frozen out.

The end result is that the final ratio of dark-matter to baryon densities is equal to the ratio of the amplitude of cosmological perturbations and the baryon-to-photon ratio, given that inflation occurred at the natural scale of the cut off of the theory.

It would be very interesting to find a mechanism which would be able to relate baryon asymmetry to the amplitude of cosmological perturbations, or other models which could connect these to the DM abundance. In addition, the possibility exists that the dark matter in our model may, in fact, be millicharged \cite{Rahman:2009rq} and couple weakly to the photon in our sector. This opens an avenue for potential direct or indirect detection. Finally, it is interesting to note that, for a sufficiently large $N$, electrons also survive and could eventually form dark atoms with the protons.

\begin{acknowledgments}
We would like to thank Gregory Gabadadze, Andrei Gruzinov, Thomas Levi, Dmitry Malyshev, Marc Manera, Michele Redi, Jonathan Roberts and Neal Weiner for useful discussions.
The work of G.D. is supported in part
by David and Lucile Packard Foundation Fellowship for Science
and Engineering, by NSF grant PHY-0245068 and by the European Commission under the ``MassTeV'' ERC Advanced Grant 226371. I.S. and A.V. are both supported
by the James Arthur Fellowship.
\end{acknowledgments}

\bibliography{NSpecies}
\bibliographystyle{JHEP}
\end{document}